\documentclass[%
reprint,
superscriptaddress,
amsmath,amssymb,
aps,
floatfix,
]{revtex4-2}
\usepackage{graphicx}
\usepackage{upgreek}
\usepackage[english]{babel}

\begin{document}
\title[Quantum detector tomography of a high dynamic-range SNSPD
]{Quantum detector tomography of a high dynamic-range superconducting nanowire single-photon detector}

\author{Timon Schapeler}\email{timon.schapeler@upb.de}
\author{Jan Philipp H\"opker}
\author{Tim J. Bartley}
\address{Mesoscopic Quantum Optics, Paderborn University, Warburger Straße 100, 33098 Paderborn, Germany}

\begin{abstract}
We demonstrate and verify quantum detector tomography of a superconducting nanowire single-photon detector (SNSPD) in a multiplexing scheme which permits measurement of up to 71000 photons per input pulse. We reconstruct the positive operator valued measure (POVM) of this device in the low photon-number regime, and use the extracted parameters to show the POVMs spanning the whole dynamic range of the device. We verify this by finding the mean photon number of a bright state. Our work shows that a reliable quantum description of large-scale SNSPD devices is possible, and should be applicable to other multiplexing configurations.
\end{abstract}

\maketitle
\section{Introduction}
Since their invention at the turn of the century~\cite{gol2001picosecond}, superconducting nanowire single-photon detectors (SNSPDs) have become ubiquitous in quantum optics experiments, due to their low noise, high speed and high efficiency~\cite{natarajan2012superconducting,marsili2013detecting,shibata2015ultimate,esmaeil2017single,korzh2020demonstration}. Although recent work has shown that some photon number resolution can be recovered directly from the output of the detector by modifying the bias circuitry~\cite{cahall2017multi,zhu2018scalable,zhu2020resolving,zou2020superconducting}, in standard operation, however, these detectors do not resolve the number of photons incident. To overcome this, SNSPDs can be multiplexed in arrays, either spatial~\cite{dauler_multi-element_2007,divochiy_superconducting_2008,marsili_physics_2009,jahanmirinejad_photon-number_2012,zhao_superconducting-nanowire_2013,rosenberg_high-speed_2013,verma_four-pixel_2014,miki_64-pixel_2014,allman_near-infrared_2015,shaw_arrays_2015,najafi_-chip_2015,chen_16-pixel_2018,tao2019high,wollman2019kilopixel} or temporal~\cite{natarajan2013quantum,tiedau2019high,jonsson2020temporal}, such that the number of pixels in the array that fire gives some information about the number of photons that were incident.

The exact nature of the response of the device to different numbers of incident photons can be formally reconstructed using quantum detector tomography~\cite{lundeen2009tomography}. This techniques provides a fully quantum mechanical description of the operation of the device, without relying on any underlying physical model of its detection mechanism. This allows one to state the probability that a specific number of photons were incident, given a particular detector outcome, know as a positive operator valued measure (POVM). 
Quantum detector tomography has been carried out for single-pixel SNDPD devices~\cite{akhlaghi2011nonlinearity,renema2012modified,endo2021quantum}, as well as temporally-~\cite{natarajan2013quantum} and spatially-~\cite{schapeler2020quantum} multiplexed SNSPDs.

At its heart, quantum detector tomography relies on a matrix inversion algorithm. The size of the matrices depends on the number of independent outcomes of the detector, as well as the dimension of the Hilbert space in which the quantum states reside. As the size of a multiplexed detector grows, not only does it exhibit more outcomes, but the Hilbert space to which it is sensitive also increases. This soon leads to a situation where the matrix inversion becomes computationally very expensive. Therefore, one of the challenges with quantum detector tomography is to develop techniques to handle large detectors, which are becoming increasingly experimentally available~\cite{miki_64-pixel_2014,allman_near-infrared_2015,wollman2019kilopixel}.

In this paper, we demonstrate tomographic reconstruction of a multiplexed detector with a Hilbert space up to $10^5$. Building on Ref.~\cite{schapeler2020quantum}, we carry out the full reconstruction of up to 11 outcomes of a multiplexed detector. This result is used to verify a model of how the POVMs behave at much higher input photon numbers, which mitigates the effect of limited computational power when reconstructing the large matrices which result from such high-dimensional reconstructions. To do so, we make use of the logarithmic time-multiplexed detector architecture, recently demonstrated with a dynamic range over 120dB~\cite{tiedau2019high}. 
We then demonstrate the utility of this technique by recovering the mean photon number of a bright coherent state.

This paper is organized as follows. We begin with a brief description of the detector under test and its applicability to recovering photon number information from bright optical states based on ensemble measurements. We then discuss detector tomography and the challenges associated with scaling to large numbers of outcomes and Hilbert space dimensions, before showing how extrapolating POVMs can be used to overcome this challenge. This is followed by the experimental methods and results, and then a verification of this approach using a bright coherent state.


\section{Achieving high dynamic range with a superconducting single-photon detector} \label{sec:highdynamic}
\begin{figure}[ht]
\centering
\includegraphics[width=0.48\textwidth]{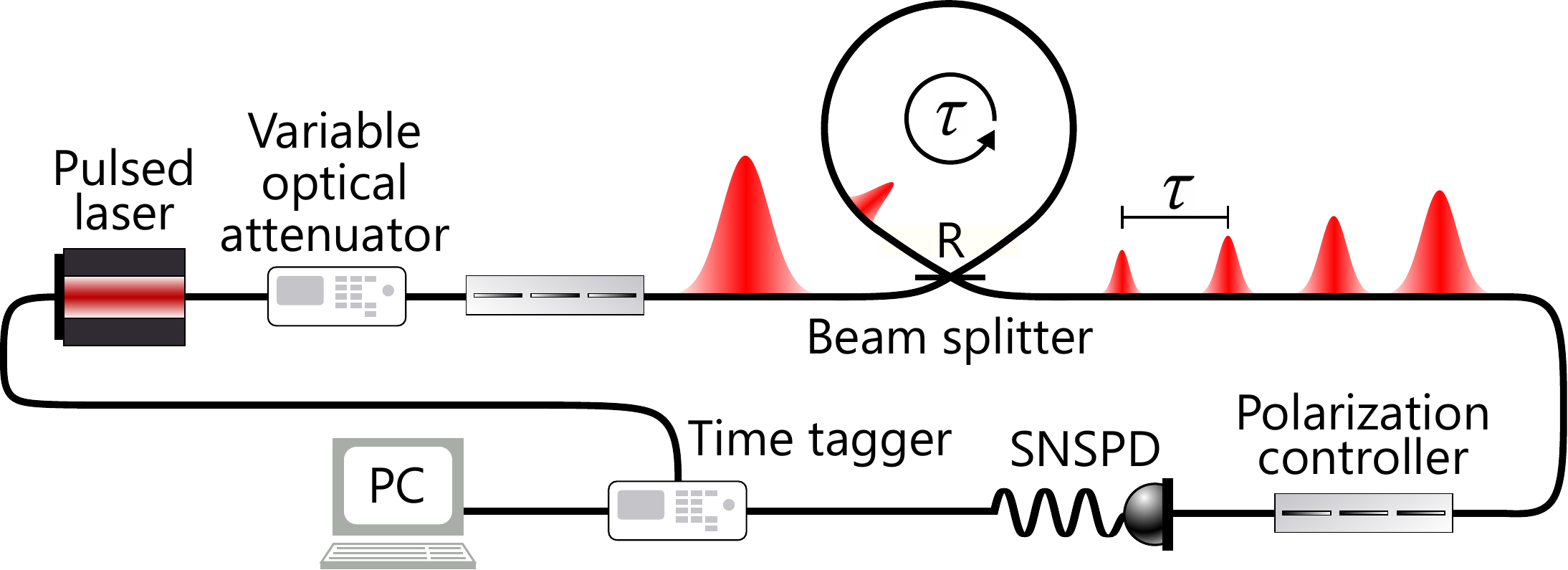}
\caption{Schematic depiction of the experimental setup and multiplexing architecture. The multiplexing is based on a beam splitter with adaptable out-coupling $R$, loop-efficiency $\eta_{\mathrm{loop}}$ and loop-length $\tau$. A 1556\,nm pulsed laser generates coherent states at a repetition rate of 15\,kHz, which can be attenuated by a computer-controlled variable optical attenuator. The first polarization controller sets the polarization for the beam splitter loop, as the components are polarization dependent. The same applies to the SNSPD, therefore, we use a second polarization controller in front of the detector. The electronic response of the SNSPD is measured by a time-tagger.}
\label{fig:Calibright}
\end{figure}

The detector upon which we will perform tomography is based on the multiplexing architecture presented in Ref.~\cite{tiedau2019high}, and shown in figure~\ref{fig:Calibright}. 
High dynamic range is achieved by splitting an input pulse into a series of output pulses with decaying intensity. The rate of intensity reduction is precisely controlled by the beam splitter and the losses inside the loop. However, these apply identically to each pulse, therefore the intensity reduction follows an exponential decay precisely~\cite{tiedau2019high}. The train of output pulses is incident on a single SNSPD, which is sensitive to each pulse if they are separated by a time greater than the recovery time of the detector, in our case the separation is $\tau=156$\,ns. In this picture, the SNSPD measures the presence or absence of photons within a time-bin. We used a standard commercial SNSPD, with nominal efficiency of 85\%, jitter $<$100\,ps and a recovery time $<$100\,ns.

For the purposes of detector tomography, we define outcomes $n$ of this detector as the number of time-bins occupied. The POVM matrix elements $\theta^{(n)}_i$ corresponding to these outcomes are thus given by the probability $p\left(n|i\right)$ of a particular outcome $n$, given a pulse containing $i$ photons.


\section{Experimental methods \& results} \label{sec:experiment}
Detector tomography is performed by subjecting a detector under test to a series of input states spanning the Hilbert space of the detector. In the case of our detector architecture, the detector cannot be saturated, since it is possible in principle for infinite time-bins to be occupied. In practice of course this is not possible, since pulses would begin to pile up, and the detector may latch. For this reason, we truncated the device after $N_{\mathrm{bin}}=10$ time-bins. Nevertheless, the size of the Hilbert space is still very large, of the order $>10^3$ photons.

In the experiment we subjected the detector under test (figure~\ref{fig:Calibright}) to a total of $D=71$ coherent states with quadratically increasing mean photon numbers $|\alpha_d|^2\approx d^2$, with $d\in[0,70]$. The coherent states are generated by a 1556\,nm laser ($g^{(2)}(0)=1.00006(17)$), which emits 9\,ps pulses at a repetition rate of 15\,kHz. The input states can be expressed in a finite matrix $\mathbf{F}$ containing Poisson distributions,
which is truncated at a maximum dimension of $M=5328$ in the photon number basis to include probability amplitudes at six standard deviation greater than the largest coherent state. For more details on the preparation and characterization of the input states see Ref.~\cite{schapeler2020quantum}.

The train of pulses, created by the beam splitter, is detected by an SNSPD (figure~\ref{fig:Calibright}), whose output pulses are measured by a time-tagger in a histogram containing $1.5\times10^5$ bins with a bin-width of 10\,ps per bin. We collect data in an ensemble measurement with a measurement time of 30\,s per coherent state. In post processing the raw time-tagger bins are integrated to find the total number of counts in each of the ten detector time-bins with a width of 2\,ns. This allows to filter out dark counts or reflections and leads to a dark-count probability of $3\times10^{-8}$ per 2\,ns time bin, which is negligibly small. 

\subsection{Defining outcomes}
The quotient of the number of counts per detector time-bin and the total number of generated laser pulses in the measurement time, describes the probabilities $p_j$ of a specific bin $j\in[1,10]$ firing. However, as mentioned in section~\ref{sec:highdynamic}, we consider an outcome $n$ of this device as the number of time-bins occupied, which leads to $N=11$ possible outcomes. Therefore, the probabilities of specific bins firing need to be transformed into probabilities $p'_n$ of $n\in[0,10]$ time-bins being occupied. This can be done using the Poisson binomial distribution, which describes a discrete probability distribution of a sum of independent Bernoulli trials, where, in general, the success probability of each trial is not identical. The probability of having $n$ time-bins occupied out of a total of $N_{\mathrm{bin}}$ can be written as the sum~\cite{wang1993number}
\begin{equation}
f(n)=\sum_{A\in F_n}\prod_{j\in A} p_j\prod_{k\in A^c}(1-p_k)~,
\end{equation}
where $F_n$ is the set of all subsets of $n$ bins that can be selected from $\{1,2,...,N_{\mathrm{bin}}\}$ and $A^c$ is the complement of $A$. As an example, the calculation of the probability of one time-bin being occupied $p'_1$, uses the set $F_1=\{\{1\},\{2\},...,\{10\}\}$. For large numbers of time-bins, this expression is expensive to calculate, as is scales with the binomial coefficient $\binom{N_{\mathrm{bin}}}{n}$. However, a closed-form expression for the Poisson binomial distribution is given by~\cite{fernandez2010closed}
\begin{equation}\label{eqn:pb}
f(n)\big|_d=\frac{1}{N_{\mathrm{bin}}+1}\sum_{l=0}^{N_{\mathrm{bin}}}C^{-ln}\prod_{j=1}^{N_{\mathrm{bin}}}\left(1+\left[C^l-1\right]p_j\big|_d\right)~,
\end{equation} 
where
\begin{equation}
    C=\exp\left(\frac{2i\pi}{N_{\mathrm{bin}}+1}\right)~,
\end{equation}
and $p_j\big|_d$ are the probabilities of a particular bin $j$ firing, given a coherent input state $d$. With this the probabilities of $n$ time-bins being occupied are $p'_n\big|_d=f(n)\big|_d$. Note that if all bins have an equal probability, {i.e.} $p_j=p\forall j$, then this distribution reduces to the binomial distribution. Finally, the outcome matrix $\mathbf{P}$ can be expressed as $P_{d,n}=p'_n\big|_d$.

\begin{figure}[tbh]
\centering
\includegraphics[width=0.48\textwidth]{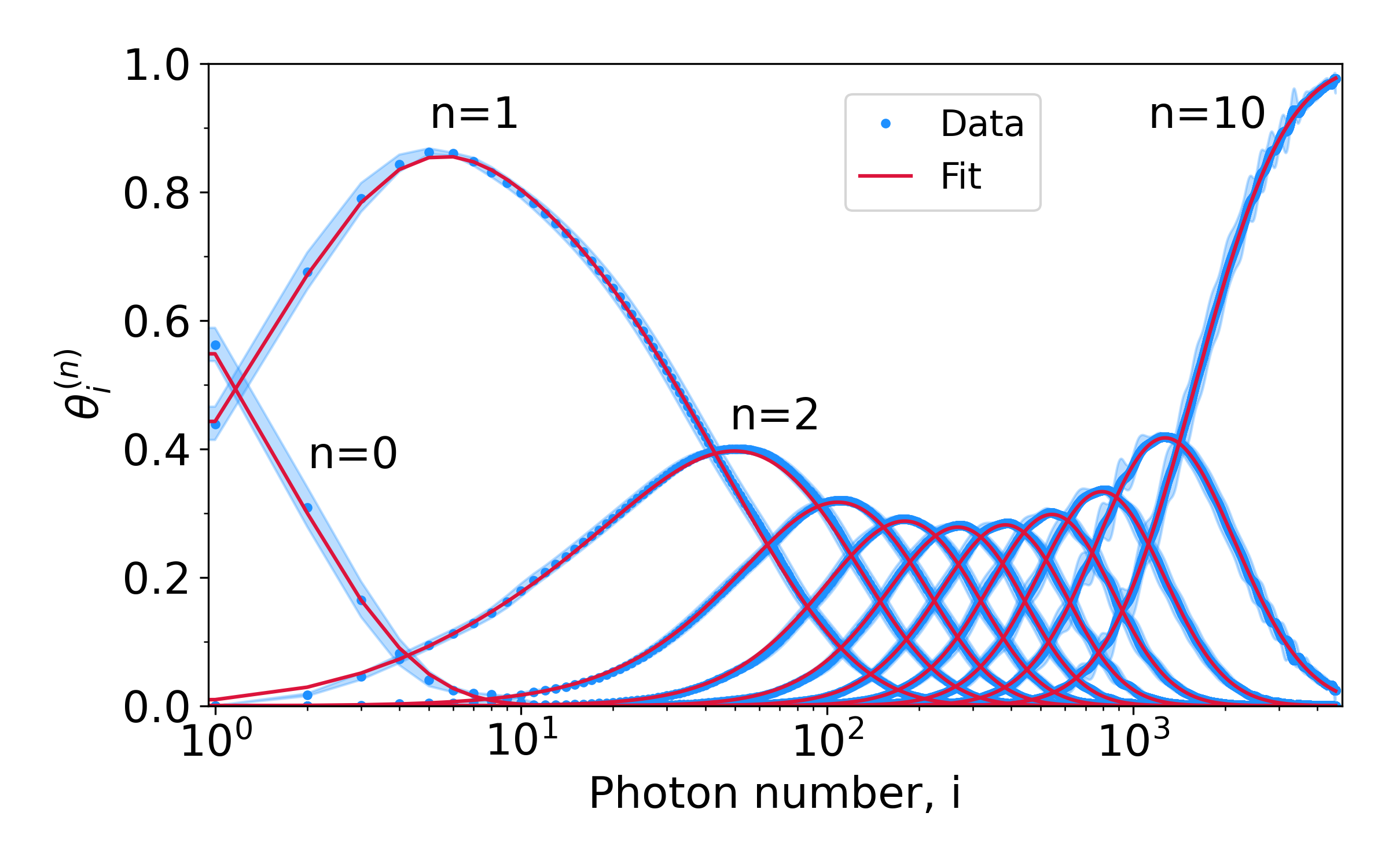}
\caption{Diagonal elements of the experimentally reconstructed POVM operators in the photon-number basis for all eleven outcomes (blue dots). With increasing detector outcomes $n$, the distributions shift to higher photon numbers $i$. The error region is based on assuming 5\% uncertainty in the amplitudes of the coherent states. The best fit (red line) using equation~\ref{eqn:pb} (compare section~\ref{sec:Extrapolation}).}
\label{fig:POVMsAndFit}
\end{figure}

\subsection{Matrix inversion}
Given the input state matrix $\mathbf{F}$ and outcome matrix $\mathbf{P}$, a matrix $\mathbf{\Pi}$ corresponding to the set of POVMs $\{\pi_n\}$ can be found by solving the optimization problem~\cite{lundeen2009tomography}
\begin{equation}
    \min\left\{\left|\left|\mathbf{P}-\mathbf{F}\mathbf{\Pi}\right|\right|_2+g\left(\mathbf{\Pi}\right)\right\}~,
\end{equation}
where $\left|\left|\cdot\right|\right|_2$ indicates the Frobenius norm~\cite{golub2013matrix} and the function
\begin{equation}
g\left(\mathbf{\Pi}\right)=\epsilon \sum_{i,n}\left(\theta^{(n)}_i-\theta^{(n)}_{i+1}\right)^2~,
\end{equation}
scaled by a factor $\epsilon$, ensures that the resulting distributions are smooth~\cite{feito2009measuring}. For more details on the reconstruction process and importance of the correct choice of the smoothing parameter $\epsilon$ can be found in Ref.~\cite{schapeler2020quantum}. The reconstructed POVM elements for all eleven outcomes are shown in figure~\ref{fig:POVMsAndFit}. The errors are calculated based on assuming 5\% uncertainty in the amplitudes of the coherent states, which stems from the uncertainty in the calibration procedure.

Using the analysis from Ref.~\cite{schapeler2020quantum} we can use the reconstructed POVMs to directly find figures of merit of the device. From this we find an overall device efficiency $\eta=0.44\pm0.03$, dark-count probability $p_{\mathrm{dark}}=(0.0+4.1)\times 10^{-7}$ and cross-talk probability $p_{\mathrm{xtalk}}=(4.6\pm2.8)\times 10^{-6}$. 

The computational effort for the reconstruction depends on the size of the POVM matrix $\mathbf{\Pi}$, {i.e.} the total number of outcomes $N$ and the Hilbert space dimension $M$. Our current tomographic reconstruction method was able to reconstruct the POVM elements of the ten time-bin detector, however, due to limited computational power (3.6\,GHz CPU with six cores and 16\,GB RAM) this seems to be the limit of this implementation. Analysis of our code indicates that it is the smoothing function which consumes the most computation time and memory, and optimizing this routine would be the starting point for improving scalability.

\begin{figure*}[tbh]
\centering
\includegraphics[width=0.88\textwidth]{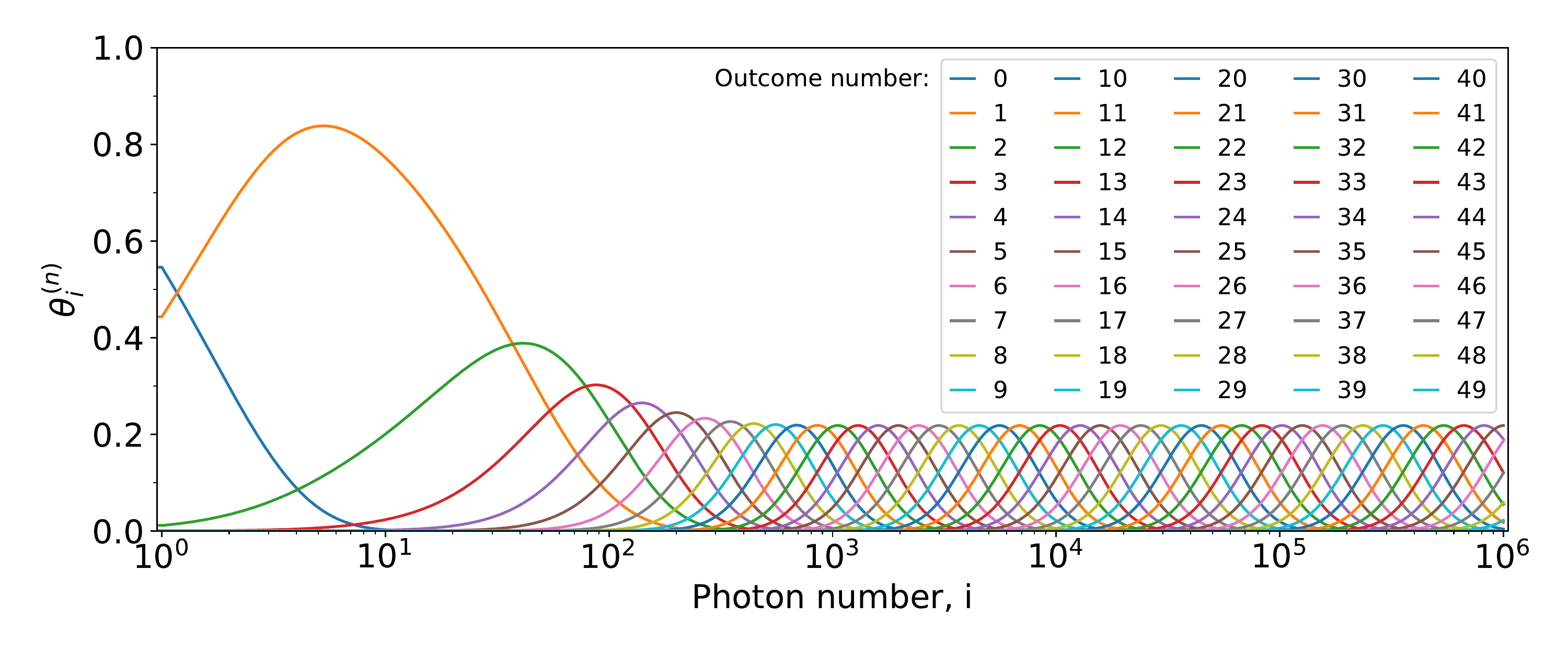}
\caption{Extrapolated POVMs using equations~\ref{eqn:pb} and~\ref{eqn:pj}, shown up to $N=50$ outcomes, corresponding to a Hilbert space dimension of $10^6$.}
\label{fig:extrapolatedPOVMs}
\end{figure*}

\subsection{Extrapolation} \label{sec:Extrapolation}
From the experimental results, it is clear that our computational power limits us to reconstructing up to eleven outcomes reliably. However, to reconstruct higher-order POVMs, a different approach is required. The crux of this approach is to use parameters calculated in the low outcome regime and extrapolate this to higher order. This deviates a little in spirit from detector tomography, since it requires some modelling of the device. However, this technique can be used to test different models of the device, since it provides a bridge between the top-down (model-free) and bottom-up (first-principles) approaches to characterizing a detector.

The first assumption about our device is that each bin has a binary result: either ``occupied'' or ``not occupied.'' Outcomes of the detector correspond to the total number of occupied time-bins. As mentioned above, this can be considered as the result of a series of independent Bernoulli trials, where, in general, each trial can have a different success probability. The general form of the probability of $n$ out of $N_{\mathrm{bin}}$ time-bins being occupied, given an input Fock state $|i\rangle$, is given by equation~\ref{eqn:pb}. 

Using this formalism, an alternative approach to reconstructing the operation of the detector is to find the probabilities $p_j(i)$ for each bin to be occupied, given the input Fock states $|i\rangle$ for $i\in[0,M]$, where $M$ is the Hilbert space dimension of the experimentally measured coherent input states. Note, however, this assumes that each bin is independent - this approach cannot account for conditional probabilities based on other bins firing, thus it cannot resolve effects like cross-talk. 

In fact, for the loop detector considered in this paper, the probabilities $p_j(i)$ can be estimated by modelling the detector. From Ref.~\cite{tiedau2019high} and neglecting dark counts (of the order $10^{-8}$ per time-bin), these probabilities are
\begin{equation}\label{eqn:pj}
    p_j\left(i\right)=
    \Bigg\{\begin{array}{ll}
         1-
         \left(1-R\eta_{\mathrm{det}}\right)^i&j=1\\
         1-
         \left[1-\frac{\left(1-R\right)^2\eta_{\mathrm{det}}}{R}
         \left(R\eta_\mathrm{loop}\right)^{j-1}\right]^i&j\geq2 
    \end{array}~,
\end{equation}
which have been adjusted to include the detection efficiency $\eta_{\mathrm{det}}$. For the specific number of time-bins $N_{\mathrm{bin}}$ of the multiplexed detector, equation~\ref{eqn:pb} depends on three parameters: the detection efficiency $\eta_{\mathrm{det}}$, the out-coupling $R$ and the loop-efficiency $\eta_\mathrm{loop}$
. The optimal set of parameters can be found by fitting the Poisson binomial approximation, for all outcomes $n\in [0,10]$ simultaneously, to the experimental data. The fit is based on a minimization of the Frobenius norm between the experimental, and fit matrix and is shown in figure~\ref{fig:POVMsAndFit} together with the experimentally reconstructed POVMs. The fit revealed the parameters $\eta_{\mathrm{det}}=0.4912\pm\left(9\times 10^{-4}\right)$, $R=0.89613\pm\left(8\times 10^{-5}\right)$ and $\eta_{\mathrm{loop}}=0.9064\pm\left(2\times 10^{-4}\right)$. 

Using these parameters and combining equations~\ref{eqn:pb} and~\ref{eqn:pj}, we can then extrapolate the POVMs out to higher order. In figure~\ref{fig:extrapolatedPOVMs} we show POVMs up to 50 outcomes, corresponding to a Hilbert space dimension of $10^6$.


\subsection{Verification using extrapolated POVMs}
To show that the reconstructed POVMs of this device from the low photon-number and low outcome regime can be used to predict the behavior of the device for higher photon-numbers and outcomes, we use the model from equation~\ref{eqn:pb} to extrapolate up to $N=120$ outcomes, which will be expressed in the matrix $\mathbf{\Pi^{\mathrm{ext}}}$. We verify this by finding the mean photon number of a bright state.

In this experiment we subject the detector to one coherent input state generated from the same laser as in section~\ref{sec:experiment} with a repetition rate of 50\,kHz. The higher repetition rate enables a measurement of the mean photon number per pulse with a power meter. The train of pulses is again detected by the SNSPD (figure~\ref{fig:Calibright}) and the output pulses are measured by a time-tagger in a histogram. We allow a total of $N_{\mathrm{bin}}=119$ time-bins, by choosing the histogram to consist of $1.87\times 10^5$ bins with a bin width of 100\,ps. The measurement time is 300\,s. By integrating over the raw bins we are able to find the probabilities of a specific bin $j\in [1,119]$ firing, which will be transformed into the probabilities $p'_n$ of $n\in [0,119]$ time-bins being occupied, using equation~\ref{eqn:pb}. These probabilities then constitute the outcome matrix $\mathbf{P}$.

Subsequently, we found the mean photon number of the bright state, by finding the coherent state which best fit the measurement data $\mathbf{P}$ and extrapolated POVMs $\mathbf{\Pi^{\mathrm{ext}}}$ by minimizing the expression
\begin{equation}
    \left|\left|\mathbf{P}-\mathbf{F^{\mathrm{fit}}}\mathbf{\Pi^{\mathrm{ext}}}\right|\right|_2~.
\end{equation}
The resulting mean photon number of $\bar{n}^{\mathrm{fit}}=71000\pm3000$ per pulse agrees with a separate measurement with a power meter of $(440\pm22)$\,pW, which results in a mean photon number of $\bar{n}^{\mathrm{pm}}=69000\pm4000$ per pulse.

\section{Conclusion}
Characterizing the response of single-photon detectors is an important task in quantum optics. Configuring these detectors to be sensitive to bright states will become increasingly important as ever brighter photonic states become available~\cite{harder2016single}. Furthermore, having a self-consistent characterization of these detectors is important wherever they are utilized. We show that a tomographic reconstruction is possible even if the dynamic range of such a detector is extremely large. This is enabled by configuring a single-photon detector, in this case an SNSPD, to respond logarithmically to the number of photons. Nevertheless, the resulting photon number can be determined with very high accuracy. While we have demonstrated this method for a time-multiplexed system, we expect a similar approach to be applicable to spatially multiplexed arrays of superconducting detectors~\cite{miki_64-pixel_2014,allman_near-infrared_2015,shaw_arrays_2015,chen_16-pixel_2018,wollman2019kilopixel}.

\section*{Acknowledgements}
We are grateful to Felix Dreher for assisting with data acquisition.
This project is supported by the German Federal Ministry of Education and Research (BMBF) under the funding program Photonics Research Germany, grant number 13N14911.


\section*{References}
\bibliography{references}

\end{document}